\documentclass[aps,prb,twocolumn,showpacs]{revtex4-1}
\usepackage{graphicx}
\usepackage{bm}
\usepackage{amsmath}
\usepackage{amssymb}
\usepackage{euscript}
\usepackage{verbatim}
\usepackage{fancyhdr}
\usepackage{wrapfig}
\usepackage{setspace}
\usepackage{xcolor}
\usepackage{amsfonts}
\usepackage{subfigure}

\newcommand{\mytitle}{Entanglement entropy and boundary renormalization group flow - exact results in the Ising universality class}

\begin{document}

\title{\mytitle}

\author{Eyal Cornfeld}
\affiliation{Raymond and Beverly Sackler School of Physics and Astronomy, Tel-Aviv University, IL-69978 Tel Aviv, Israel}

\author{Eran Sela}
\affiliation{Raymond and Beverly Sackler School of Physics and Astronomy, Tel-Aviv University, IL-69978 Tel Aviv, Israel}


\begin{abstract}
The entanglement entropy in one dimensional critical systems with boundaries has been associated with the noninteger ground state degeneracy. This quantity, being a characteristic of boundary fixed points, decreases under renormalization group flow,
as predicted by the g-theorem. 
Here, using conformal field theory methods, we exactly calculate the entanglement entropy in the boundary Ising universality class. Our expression can be separated into the well known bulk term
and a boundary entanglement term, displaying a universal flow between two boundary conditions, in accordance with the g-theorem. These results are obtained within the replica trick approach, where we show that the associated twist field, a central object generating the geometry  of an $n$-sheeted Riemann surface, can be bosonized, giving simple analytic access to multiple quantities of interest. We argue that our result applies to other models falling into the same universality class. This includes the vicinity of the quantum critical point of the two-channel Kondo model, allowing to track in real space the presence of a region containing one half of a qubit with entropy $\frac{1}{2} \log (2)$, associated with a free local Majorana fermion.
\end{abstract}

\maketitle

\section{Introduction}
Boundary critical phenomena~\cite{cardy1984conformal} have found innumerate applications in condensed matter systems and string theory. A key result in 1D critical systems, 
known as the g-theorem~\cite{affleck1991universal}, states that upon renormalization group (RG) flow, the noninteger ground state degeneracy, $g$, as well as the entropy associated with the boundary, $\log(g)$, necessarily decrease, signaling a quenching of the boundary's degrees of freedom. 

Entanglement entropy (EE), on the other hand, has recently found multiple connections in high energy physics and black holes~\cite{ryu2006holographic,azeyanagi2008holographic,fujita2011aspects,jensen2013holography}, as well as in condensed matter, relating to quantum phase transitions, topological phases, and developments of numerical algorithms; for reviews see~\cite{verstraete2008matrix,amico2008entanglement,horodecki2009quantum,calabrese2009entanglement,laflorencie2016quantum}. Moreover, first measurements of entanglement have become possible using twin many-body systems~\cite{islam2015measuring}. 

In critical 1D systems with a boundary, the universal noninteger degeneracy has been identified by Calabrese and Cardy, as a subleading term in the entanglement entropy~\cite{calabrese2004entanglement},
\begin{equation}
\label{CCformula}
S_A(\ell) = \frac{c}{6} \log \frac{2 \ell }{a}+ \log (g)+c_1'.
\end{equation} 
Here, $c$ is the central charge, $\ell$ defines the bipartition of the system, $a$ is a short distance scale and $c'_1$ is a nonuniversal constant. In general, there are boundary perturbations which could be relevant or irrelevant. Then, $\log (g)$ is no longer a constant but rather flows under RG~\cite{eriksson2011corrections,PhysRevB.84.041107,castro2009bi},
\begin{equation}
\log (g) \to \log (g(\ell)).
\end{equation}
For a relevant perturbation, $\log (g)$ decreases upon increasing $\ell$ according to the g-theorem between two known universal values; see Ref.~[\onlinecite{Casini2016}] for a proof of the g-theorem in the context of entanglement and Refs.~[\onlinecite{erdmenger2013holographic},\onlinecite{erdmenger2016entanglement}] for a holographic view in the context of the Kondo effect. 
Several numerical efforts have been carried in order to identify this noninteger ground state degeneracy in the subleading term of the EE, and its RG flow. This was accomplished in various models, including Kondo models~\cite{sorensen2007impurity,alkurtass2016entanglement} and the boundary Ising chain~\cite{zhou2006entanglement}; for a review see Ref.~[\onlinecite{affleck2009entanglement}].

\begin{figure}[t]
%\centering	
\includegraphics[width=1\linewidth]{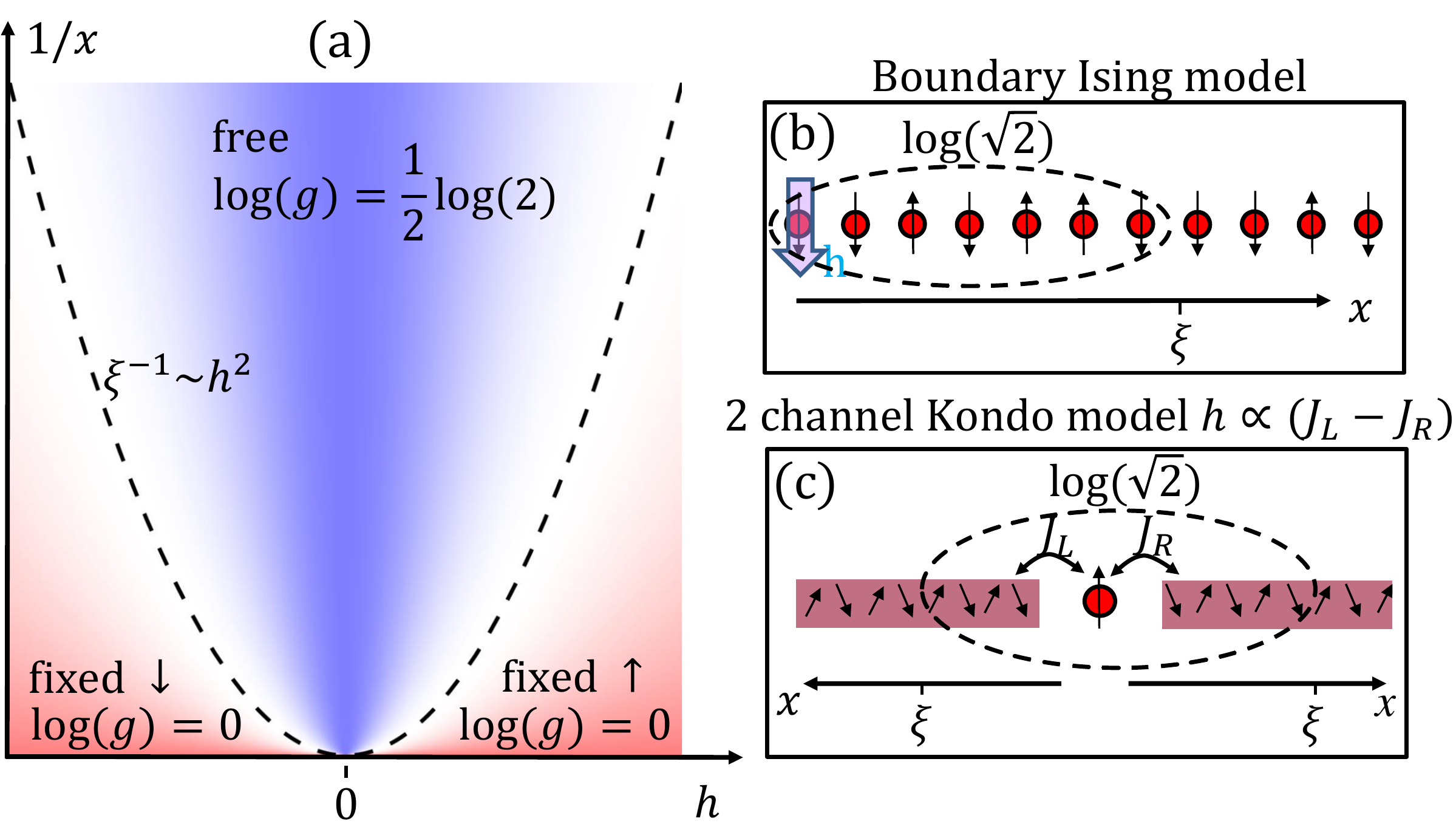}
\caption{(a) Quantum critical point of the boundary Ising universality class. A crossover length scale, $\xi \thicksim h^{-2}$, separates the free boundary condition (BC) with finite entropy at short distances from fixed BC with quenched entropy at long distances. Specific model examples include (b) semi-infinite critical transverse field Ising chain, where $h$ corresponds to a magnetic field applied at the boundary spin, or (c) two-channel Kondo model where an impurity spin couples to two baths of conduction electrons with ${h \propto J_L - J_R}$. \label{fig:3parts}} 	 	
\end{figure}

Of central interest is the Ising universality class of boundary critical phenomena, displaying a quenching of a ground state degeneracy of $\frac{1}{2} \log 2$ at the boundary, \emph{i.e} that of half a qubit, with applications to systems hosting Majorana fermions, spin systems near phase transitions, or quantum impurity models, see Fig.~\ref{fig:3parts}.
A number of fruitful techniques had been applied to the EE in this theory, either based on its simple free fermion structure~\cite{Casini2016}, or via the form-factor approach~\cite{cardy2008form,castro2009bi,saleur2013entanglement,vasseur2014universal,Casini2016}. With an eye on possible generalizations to other theories, finding additional convenient techniques for the Ising model may be highly valuable.

In this paper we present a new method based on conformal field theory, which allows us to obtain exact analytic results describing the entire flow in real space of the entanglement entropy in the boundary Ising universality class. As we show, this result is applicable in the context of two-channel~\cite{lee2015macroscopic,alkurtass2016entanglement} or two-impurity Kondo systems~\cite{bayat2010negativity}, and allows the real space identification of the two-channel Kondo screening cloud~\cite{affleck2001detecting,sorensen2007impurity,mitchell2011real,park2013directly}, which hosts a Majorana fermion half-qubit degree of freedom.

Our exact analytic result nicely fits existing detailed numerical results for the boundary Ising chain~\cite{zhou2006entanglement}. It also agrees with a result for an equivalent model  solved in Ref.~[\onlinecite{Casini2016}] based solely on free fermion techniques. Our fully analytic approach is facilitated via a bosonization scheme for the so called twist-field, an object in the field theory that realizes the $n$-sheeted Riemann geometry and whose correlation function encodes entanglement properties~\cite{calabrese2004entanglement,cardy2008form}. Our method is expected to allow additional calculations in the boundary Ising universality class, such as negativity~\cite{calabrese2012entanglement,calabrese2013entanglement,calabrese2013entanglement1,HOOGEVEEN2015,Calabrese2015Finite} or entanglement of multiple intervals~\cite{calabrese2011entanglement,cardy2013some,Rajabpour2012}.

\section{Entanglement entropy}
\begin{figure*}[t]
 \centering	
 \includegraphics[width=1\linewidth]{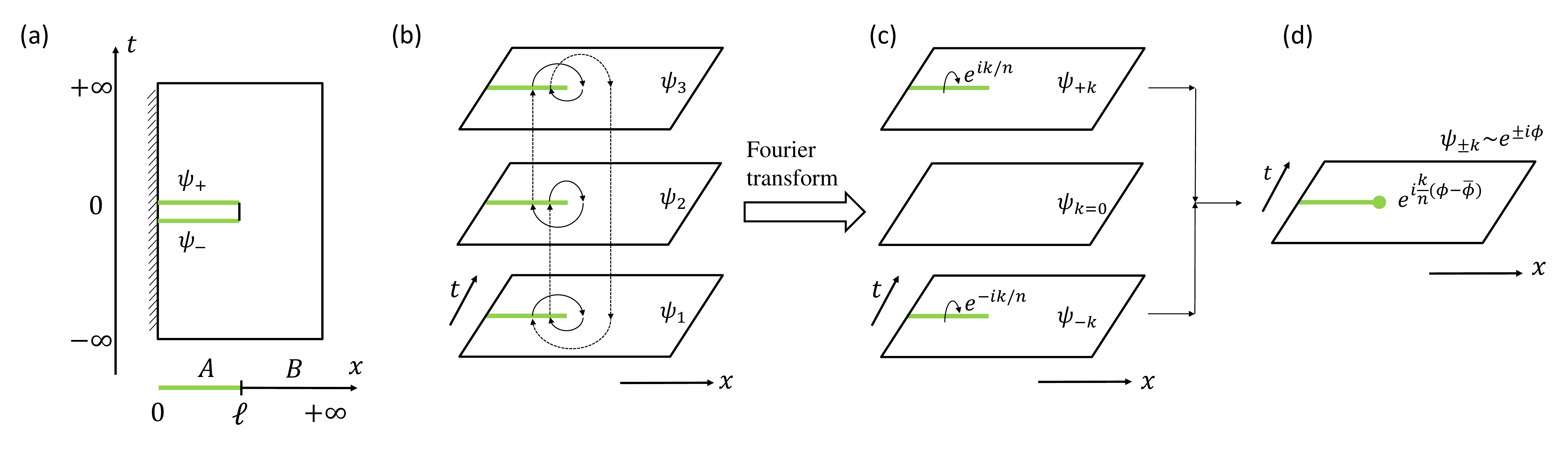}
 \caption{(a) A semi-infinite quantum chain ${x \ge 0}$ maps into a field theory in 1+1 dimensions with a line boundary parametrized by $t$. Performing a cut at ${t=0}$ for $x \le \ell$, and introducing $n$ replicas of the system allows one to construct (b) the $n$-sheeted Riemann surface, $\mathcal{R}_n$. Computing the partition function on this geometry provides $\mathrm{tr} \rho_A^n$, the essential ingredient for the EE. (c) Upon Fourier transforming over the sheet index, one obtains $n$ new semi-infinite planes. When a field crosses the cut it just picks up a phase of $e^{2\pi ik/n}$. (d) For the Ising model, the theory on each plane is that of a free Majorana fermion. One can pair up the $\pm k$ Majorana sectors into a single plane and construct one complex fermion $\psi_k$ which can be bosonized along with the twist field ${\mathcal{T}_k \thicksim e^{i \frac{k}{n} (\phi - \bar{\phi})}}$. \label{fig:3sheet}}
 \end{figure*}

We briefly review the connection between entanglement entropy and geometry~\cite{calabrese2004entanglement,cardy2008form}. The entanglement entropy, ${S = - \mathrm{tr} \rho_A \log \rho_A}$, of subsystem $A$ with a reduced density matrix ${\rho_A = \mathrm{tr}_B \rho}$, where $\rho$ is the full density matrix, is a limit case of the R\'{e}nyi entropy ${S_n=\frac{1}{1-n} \log \mathrm{tr} \rho_A^n}$. A physical way to interpret this relation, ${S =\lim_{n \to 1} S_n}$, is via the replica trick explained herein. Consider a semi-infinite 1D system as in Fig.~\ref{fig:3sheet}(a). While $\mathrm{tr} \rho$ is the partition function that can be expressed as a path integral over a semi-infintie plane,   ${\mathrm{tr}} \rho_A^n$ is the partition function of the same theory on the geometry constructed from $n$ copies of the semi-infinite plane, cut along the segment $t=0$, ${0 \le x \le \ell}$, and glued together into an $n$-sheeted Riemann surface structure; see Fig.~\ref{fig:3sheet}(b).

Refs.~[\onlinecite{calabrese2004entanglement},\onlinecite{cardy2008form}] introduced a local twist field, $\mathcal{T}$, such that any correlation function on the $n$-sheeted Riemann surface, $\mathcal{R}_n$, can be computed on $n$ decoupled half-planes, $\mathcal{R}^n$, as 
\begin{equation}
\label{nsheet}
\langle \mathcal{O} \rangle_{\mathcal{R}_n} = \frac{\langle  \mathcal{O} ~ \mathcal{T}(w,\bar{w}) \rangle_{\mathcal{R}^n}}{\langle \mathcal{T}(w,\bar{w})  \rangle_{\mathcal{R}^n}}.
\end{equation}
Here, ${(w,\bar{w}) = (i \ell , - i \ell)}$ is the source of the branch cut in complex coordinates ${(z , \bar{z}
)= (t + i x,t - i x)}$. 
The computation of ${\mathrm{tr}} \rho_A^n$
boils down to that of the one-point function of the twist field generating the geometry of the $n$ copies, ${\mathrm{tr} \rho_A^n\propto\langle \mathcal{T}(w,\bar{w}) \rangle}$. To obtain the EE, this replica trick only becomes useful provided that analytic continuation to noninteger $n$ can be taken. For a conformal field theory in 1+1 dimensions, the $n$-sheeted Riemann surface is related to the complex plane by a conformal transformation, see below. These observations lead to the conclusion that the twist field behaves as a primary field with a scaling dimension~\cite{calabrese2004entanglement,cardy2008form,Knizhnik1987,dixon1987conformal} ${h_\mathcal{T}=\frac{c}{24}\left( n - 1/n \right)}$, allowing the elegant derivation of the EE scaling, Eq.~(\ref{CCformula}), in conformal invariant systems.

\section{Ising boundary universality class}
The Ising model at the critical temperature admits a Lagrangian description in terms of a ${c=1/2}$ conformal field theory (CFT) of free massless Majorana Fermi field $(\psi, \bar{\psi})$. In the presence of a boundary $\mathcal{B}$, which is taken to be the line ${x=0}$ in Fig.~\ref{fig:3sheet}, the action is~\cite{ghoshal1994boundary} 
\begin{multline}
\label{Sh} 
S_h = \frac{1}{2 \pi} \int_\mathcal{D} d^2 z\bigg[\psi \partial_{\bar{z}} \psi +\bar{\psi} \partial_z \bar{\psi}\bigg]  \\
 + \int_\mathcal{B} dt \bigg[- \frac{i}{4 \pi} \psi \bar{\psi}+\frac{1}{2} a \dot{a}\bigg]
+ ih\int_\mathcal{B} dt \bigg[
a\cdot(\psi+\bar{\psi})\bigg],
\end{multline}
where ${z = t+i x}$ and ${\partial \mathcal{D} =  \mathcal{B}}$.  Here, $a(t)$ is a local Majorana fermion ${\langle a(t) a(t') 
\rangle = \frac{1}{2} {\mathrm{sign}}(t-t')}$ and $h$ is a magnetic field applied at the boundary which represents a relevant boundary perturbation leading to an RG flow. The fermion correlation function, ${\langle \psi(z) \psi(z') 
\rangle = \frac{1}{z-z'}}$, implies that the scaling dimension of this perturbation is $1/2$. In turn, finite $h$ leads to a finite correlation length $\xi \thicksim h^{-2}$. 

The crossover phase diagram of this model is shown in Fig.~\ref{fig:3parts}(a) for small $h$. The $y$ axis is the inverse distance from the boundary (it can also be thought of as temperature). At small length scales, $1/x \gg 1/\xi$, the perturbation is weak and there is a ground state degeneracy of half a qubit, ${\log (g)=\frac{1}{2} \log (2)}$, associated with the Majorana fermion $a$.  At large length scales, $1/x \ll 1/\xi$, this local Majorana fermion hybridizes with the Majorana field $\psi$, hence quenching the boundary degeneracy, ${\log (g)=0}$.

This field theory describes the long distance physics of a class of problems falling into the same universality class. The most directly connected lattice model is the boundary-Ising chain; see Fig.~\ref{fig:3parts}(b),
\begin{equation}
\label{bim}
H =  - \sum_{j=0}^{\infty} S_j^z S_{j+1}^z + \tfrac{1}{2}\cdot S_j^x +h_b \cdot S_0^z.
\end{equation}
The bulk transverse field is tuned to its critical value, $1/2$. A weak longitudinal
field, $h_b$, is applied at the boundary spin, ${j = 0}$. The boundary field, $h_b$, maps to the relevant perturbation, $h$, in the field theory. This implies an RG flow between the two
fixed points corresponding to a free boundary condition (BC), ${h_b=0}$, and a fixed BC, ${h_b = \pm \infty}$. The noninteger ground state degeneracies associated with these BCs, given by ${g = \sqrt{2}}$ and ${g = 1}$, respectively, were identified in numerical studies of the EE~\cite{affleck2009entanglement,zhou2006entanglement}. 

As the field theory implies, the entropy of $\frac{1}{2} \log (2)$ at the free boundary condition is connected with unpaired Majorana fermions which recently attracted considerable attention. Such localized Majorana fermion zero energy states also arise in the two-channel Kondo model~\cite{emery1992mapping}, where an impurity spin-$1/2$ antiferromagnetically couples to two channels of electrons; see Fig.~\ref{fig:3parts}(c). This leads to frustration, and hence to a partial screening of the impurity entropy from $ \log (2)$, corresponding to the decoupled spin-$1/2$ impurity, down to $\frac{1}{2} \log(2)$. This partial quenching of the boundary entropy has been numerically identified in the EE~\cite{lee2015macroscopic,alkurtass2016entanglement}. The applicability of the field theory $S_h$ to the two channel Kondo model has been identified in Ref.~[\onlinecite{sela2011exact}], wherein, $h$ corresponds to one of few perturbations of the model~\cite{mitchell2012universal}, \emph{e.g.} channel anisotropy, where one channel couples more strongly to the impurity spin.  At low temperatures, the weakly coupled channel fully decouples, and the strongly coupled channel of spin-$1/2$ electrons fully screens the impurity spin into a spin singlet state with ${\log (g)=0}$. 

In fact, many different quantum impurity models, such as the two-impurity Kondo model~\cite{affleck1995conformal,sela2009nonequilibrium1,sela2009nonequilibrium,mitchell2012two}, map to the same field theory and share the same quenching of a half-qubit entropy.

\section{Bosonization of the twist field}
One can trade the complicated n-sheeted Riemann geometry by a trivial semi-infinite plane geometry, in which each field of the theory, such as the fermion field $\psi$, is replaced by an $n$-component field ${(\psi_1 , \psi_2,\ldots,\psi_n)^{T}}$; upon crossing the cut, ${0 \le x \le \ell}$, the fields satisfy a boundary condition ${(\psi_{+})_j = (\psi_{-})_{j-1}}$; see Fig.~\ref{fig:3sheet}(b). In other words, the $n$-component field gets multiplied by a twist matrix $\mathrm{T}$ given by ${\mathrm{T}_{ij} = \delta_{i-j+1}}$ for even $n$ and by a similar expression for odd $n$~\cite{casini2005entanglement}. This matrix can be diagonalized in Fourier basis, ${\psi_k = \frac{1}{\sqrt{n}} \sum_j e^{2\pi i j (k-n/2)/n} \psi_j}$, with ${\psi_k^*(z)=\psi_{-k}^{\phantom{*}}(z)}$, yielding ${\mathrm{T} \psi_k =e^{2\pi i k /n} \psi_k}$. The range of $k$ values is given by
\begin{equation}
\label{range}
k=-(n-1)/2,\ldots,(n-1)/2.
\end{equation}
For even $n$, one may decompose the theory into $n/2$ \emph{complex} fermions $(\psi_k^{\phantom{*}}, \psi^*_k)$ with ${k=1/2,\ldots,(n-1)/2}$. For odd $n$, there is an unpaired Majorana fermion $\psi_0$ and $(n-1)/2$ complex fermions ${k=1,\ldots,(n-1)/2}$; see Fig.~\ref{fig:3sheet}(c). Upon crossing the cut, the $k$-th fermion acquires a phase $e^{2\pi i k /n}$, while the ${k=0}$ unpaired Majorana fermion $\psi_0$ for odd $n$ is insensitive to the cut. 

The nontrivial phase factor, $e^{2\pi i k/n}$, picked up by the fermions upon crossing the cut, originates from the twist field located at the end of the cut, ${x = \ell}$, as in Eq.~(\ref{nsheet}); see Fig.~\ref{fig:3sheet}(d). 
To extract the interplay between the twist field and the complex fermions, in terms of their bulk operator product expansion (OPE), we consider ${\mathcal{O} = \psi_k^*(z) \psi_{k}^{\phantom{*}}(z')}$ in Eq.~(\ref{nsheet}) and apply the conformal transformation ${\xi(z) = ( \frac{z-w}{z-\bar{w}})^{1/n}}$, yielding the correlation function,
\begin{equation}
\label{X}
\frac{\langle \psi_{k}^*(z)\psi_{k}^{\phantom{*}}(z') \mathcal{T}(w,\bar{w})  \rangle}{\langle \mathcal{T}(w,\bar{w}) \rangle} =\frac{1}{z-z'}\left(\frac{(z-w)(z'-\bar{w})}{(z-\bar{w})(z'-w)} \right)^{k/n}.
\end{equation}
For a calculation see Appendix~\ref{app:3pf}. Interesting insights follow from this result. First, upon taking $\psi_k(z)$ around the twist field at $w$, the correct factor of $e^{2\pi ik/n}$ is recovered. Second, from the diagonal operation of the twist on the paired Fourier transformed complex fermion fields, $(\psi_k^{\phantom{*}}, \psi^*_k)$, one may infer a similar factorization of the twist field
\begin{equation}
\label{factori}
\mathcal{T}(w,\bar{w})= \prod_{k\ge 0} \mathcal{T}_k(w,\bar{w}),
\end{equation}
where $\mathcal{T}_k$ is the part of the twist field acting on the $\pm k $ Majorana components. While we know the total scaling dimension ${h_\mathcal{T} = \sum_{k \ge 0} h_k}$, what is the scaling dimension of each component $ \mathcal{T}_k$?
We can extract it from 
\begin{equation}
\frac{ \langle T_k(z)  \mathcal{T}(w,\bar{w})  \rangle}{\langle \mathcal{T}(w,\bar{w}) \rangle}
= h_k \frac{(w-\bar{w})^2}{(z-w)^2(z-\bar{w})^2}.
\end{equation}
The stress-energy tensor $T_k$ of the $\pm k$ sectors, being two Ising theories, is given by that of a complex fermion theory, ${T_k = -\frac{1}{2} (\psi^*_k \partial_z \psi_k^{\phantom{*}} - \partial_z \psi^*_k \psi_k^{\phantom{*}})}$, obtained from Eq.~(\ref{X}) by applying ${\frac{1}{2 \pi i}\oint \frac{dz'}{z'-z} [-\frac{1}{2}(\partial_{z'}-\partial_{z})]}$, yielding
\begin{equation}
\label{hk}
h_k=\frac{k^2}{2 n^2}.
\end{equation}
For odd $n$, the ${k=0}$ sector has ${h_k=0}$, i.e. this component of the twist field acts as the identity field and hence does not contribute to any calculation. For all $n$, Eq.~(\ref{hk}) sums up correctly to the total scaling dimension of the twist field ${h_\mathcal{T} = \sum_{k>0}^{(n-1)/2} h_k = \frac{1}{48}(n-1/n)}$.

Putting together the monodromy that follows from Eq.~(\ref{X}) as well as the scaling dimension $h_k$ of the twist field Eq.~(\ref{hk}), we deduce that the twist field admits a bosonization formula. By introducing a boson field $\phi$ for each  sector $\pm k$ and writing the complex fermion field as ${\psi_k = e^{i \phi}}$ and ${\bar{\psi}_k= e^{i \bar{\phi}}}$ we infer    
\begin{equation}
\label{bosoni}
\mathcal{T}_k(w,\bar{w})=e^{i \frac{k}{n}(\phi(w) - \bar{\phi}(\bar{w}))}.
\end{equation} 
By introducing the vertex operators ${V_\alpha(w) = e^{i \alpha \phi(w)}}$, ${\bar{V}_\alpha(\bar{w}) = e^{i \alpha \bar{\phi}(\bar{w})}}$, the Fermi and twist fields take the form, ${\psi = V_1}$, ${\bar{\psi} = \bar{V}_1}$, and ${\mathcal{T}_k(w,\bar{w})=V_{k/n}(w) \bar{V}_{-k/n}(\bar{w})}$.
	   
Equations~(\ref{factori}) and (\ref{bosoni})  are the central result of this section. This procedure of pairing of $\pm k$ Majorana sectors followed by bosonization, is related to the well known two-copy bosonization of the Ising CFT which allows the calculation of all its critical correlations~\cite{di1987critical}. The Ising order parameter, $\sigma$, also admits a bosonization formula ${\sigma^2 \propto V_{\frac{1}{2}} \bar{V}_{-\frac{1}{2}}+V_{-\frac{1}{2}} \bar{V}_{\frac{1}{2}}}$. Hence, the spin field and the twist field have a similar structure in terms of vertex operators $V_\alpha$. However, note that ${\alpha=1/2}$ is not contained in the set of $k/n$ in Eq.~(\ref{range}). Thus, the spin field is equivalent to a continuation of the twist field, $\mathcal{T}_k$, to ${k = n/2}$. We note that similar bosonization techniques were used in Refs.~[\onlinecite{cardy2008form},\onlinecite{Belin2013}].

We emphasize that while CFTs are well understood objects, $n$-copies of a  CFT endowed with the twist field form a more complicated object known as an orbifold~\cite{dixon1987conformal}. The OPEs of the twist fields with other fields~\cite{Headrick2010} have been extensively explored in many circumstances~\cite{calabrese2011entanglement,Rajabpour2012,HOOGEVEEN2015,Calabrese2015Finite,BIANCHINI2015}, but nevertheless remain elusive in general. The bosonization procedure which is known for orbifolds of free theories~\cite{dixon1987conformal}, as applied here in the context of entanglement, leads to a number of applications. 
For example, one may apply it to compute the entanglement of multiple intervals, corresponding to an insertion of multiple twist fields, which has been carried out using other methods~\cite{calabrese2011entanglement,cardy2013some,Rajabpour2012}. Hereafter, we demonstrate the power of this method for the calculation of the entanglement entropy in the presence of a boundary with a magnetic field.

\section{Differential equation for the R\'{e}nyi entropy}
\emph{Our goal is to obtain a differential equation for the one-point function ${\langle \mathcal{T}(w=i\ell,\bar{w}=-i\ell) \rangle_\lambda}$ in the presence of the boundary field, $h$, which breaks boundary conformal invariance.}  Knowledge of this expectation value, analytically in $n$, will give the desired entanglement entropy,
\begin{equation}
S_n(\ell)_\lambda=\tfrac{1}{1-n}\ln\mathrm{tr}\rho_A^n=\tfrac{1}{1-n}\ln\langle \mathcal{T}(i\ell,-i\ell) \rangle_\lambda + const.
\end{equation}
Here, ${\xi^{-1}=\lambda=4\pi h^2}$ is the inverse correlation length.
\subsection{Method of Chatterjee and Zamolodchikov}
In order to compute the one-point function of the twist field in the boundary Ising model, we use the method of Chatterjee and Zamolodchikov~(CZ)~\cite{chatterjee1994local}. The method is based on the property that the boundary condition for any $h$ is simple in terms of the Fermi fields ($\psi,\bar{\psi})$. Consider first the conformal cases ${h=0}$ or ${h = \infty}$. CFT of the Ising model in the presence of boundary (along with
more general conformal field theories) had been studied in Refs.~[\onlinecite{cardy1989boundary},\onlinecite{cardy1991bulk}] and it was shown that there are two conformal invariant boundary conditions, 
\begin{equation}
[\psi - \bar{\psi}]_\mathcal{B}=0,~\mathrm{(free)},~~~[\psi + \bar{\psi}]_\mathcal{B}=0,~\mathrm{(fixed)}.
\end{equation}
Such boundary conditions allow one to move from the semi-infinite half plane $x>0$ to the infinite plane, and regard $\bar{\psi}$ as the analytic continuation of $\pm \psi$ at the $x<0$ half plane; the $\pm$ signs correspond to the two boundary conditions. Next, consider the action Eq.~(\ref{Sh}) at finite $h$. From the equations of motion, one finds~\cite{chatterjee1994local}
\begin{equation}
(\tfrac{d}{d t} + i \lambda) \psi(t) = (\tfrac{d}{d t} - i \lambda) \bar{\psi}(t),~~~~~~\lambda =4 \pi  h^2.
\end{equation}
Equivalently, ${(\partial_z + i \lambda) \psi(z) = (\partial_{\bar{z}} - i \lambda) \bar{\psi}(\bar{z})}$. This form of the boundary condition makes it explicit that the fields
\begin{equation}
\label{eqchi}
\chi(z) = (\partial_z + i \lambda) \psi(z),~~~~\bar{\chi}(\bar{z}) = (\partial_{\bar{z}} - i \lambda) \bar{\psi}(\bar{z})
\end{equation}
enjoy the desired property that $\bar{\chi}(\bar{z})$ coincides with the analytic continuation of $\chi(z)$ to the $x<0$ half-plane.

CZ used this property to derive differential equations for the magnetization, $\langle \sigma(x) \rangle$, at distance $x$ from the boundary. This requires to identify an OPE of the fermion field with some other field that results in the desired operator, $\sigma$. Indeed, one can use the known OPE, ${\psi\times\mu=\sigma}$,
with known coefficients,
\begin{equation}
\label{psiOPE}
\psi(z) \mu(w,\bar{w}) =\frac{e^{-i\pi/4}}{\sqrt{2}}(z-w)^{-1/2}\sigma(w,\bar{w})+\ldots.
\end{equation}
It is subsequently straightforward to obtain the coefficients in the OPEs, ${\chi\times\mu=\sigma}$ and ${\bar\chi\times\mu=\sigma}$.
We emphasize that these OPE coefficients are bulk properties, insensitive to the boundary.
	   	  	   	  
Finally, CZ introduced the following \emph{meromorphic function}: Consider the expectation values, $\langle \chi(z) \mu(w,\bar{w}) \rangle_\lambda
$ or $\langle \bar\chi(\bar{z}) \mu(w,\bar{w}) \rangle_\lambda
$, in the presence of the boundary field. Using the boundary condition  ${[\chi - \bar{\chi}]_\mathcal{B}=0}$, one can extend to the full $z$-plane and view this correlation function as an analytic function with two branch-cut points at ${z = w}$ and ${z = \bar{w}}$. Taking into account Eq.~(\ref{psiOPE}) and the asymptotic behavior
\begin{equation}
\chi(z) \thicksim z^{-1},~~~ z \to \infty,
\end{equation}
one can write
\begin{multline}
\label{meroAB}
\langle \chi(z) \mu(w,\bar{w}) \rangle =\frac{1}{(z-w)^{1/2}(z-\bar{w})^{1/2}} \\
\times\left( \frac{A(w,\bar{w})}{z-w} +\frac{\bar{A}(w,\bar{w})}{z-w} +B(w,\bar{w}) \right).
\end{multline}
Using the explicitly known OPE coefficients of $\chi\times\mu$, one can linearly relate the functions $A,\bar{A}, B$, to $\langle  \sigma(w, \bar{w}) \rangle$ and its derivatives, and attain a differential equation that fully determines the magnetization~\cite{chatterjee1994local}. 
	   	  	   	 
Since our bosonization scheme allows the tracking of the OPE of $\mathcal{T}$ with all other fields in the theory using the known OPE of vertex operators, we can now employ the above CZ procedure for the twist field.

\subsection{Generalized CZ method for the twist field}\label{GCZ}
We use the decomposition of the twist field Eq.~(\ref{factori})  and the bosonization formula Eq.~(\ref{bosoni}). For ${\lambda=0}$ or ${\lambda = \infty}$, i.e. conformal invariant boundary conditions, the correlation of the vertex operators $V_{k/n}$ and $\bar{V}_{k/n}$ depends solely on their holomorphic scaling dimension, ${h_k=k^2/(2n^2)}$, giving a powerlaw decay
\begin{equation}
\langle \mathcal{T}_k(i \ell,-i \ell)  \rangle_{\lambda=0,\infty} \propto \frac{1}{(2\ell)^{k^2/n^2}},
\end{equation}
so that
\begin{equation}
\prod_{k>0}^{(n-1)/2}\!\!\! \langle \mathcal{T}_k(i \ell,-i \ell)  \rangle_{\lambda=0,\infty} \propto \frac{1}{(2\ell)^{(n-1/n)/24}},
\end{equation}
reproducing the formula, ${S_1(\ell)=\frac{1}{12} \log \frac{2 \ell}{a}+const},$ for $\lambda=0,\infty$.

We wish to find the non-logarithmic corrections to the entanglement entropy, $S_n$; hence, it will be productive to introduce the normalized dimensionless twist, $\mathfrak{t}_{k/n}$, and the boundary entropy, $\mathfrak{s}_n$,
\begin{gather}
\label{fdef}\mathfrak{t}_{k/n}(\lambda\ell)=\frac{\langle\mathcal{T}(i\ell,-i\ell)\rangle_\lambda}{\langle\mathcal{T}(i\ell,-i\ell)\rangle_{\lambda=\infty}},\\
\label{BEdef}\mathfrak{s}_n(\ell) = S_n(\ell)_\lambda- S_n(\ell)_{\lambda=\infty}= \frac{1}{1-n}\ln\!\!\!\prod_{k>0}^{(n-1)/2}\!\!\!\mathfrak{t}_{k/n}(\lambda\ell),
\end{gather}
such that ${\mathfrak{t}_{k/n}|_{\lambda\ell\to\infty}=1}$. This boundary entropy directly relates to the groundstate degeneracy, ${\ln(g(\ell))=\mathfrak{s}_1(\ell)}$, and trivially satisfies ${\ln(g)|_{\lambda\to\infty}=0}$.

To treat the non-conformal invariant case with finite $\lambda$, we return to the CZ method. We make use of the known OPE of vertex operators in the free-boson theory ${V_\alpha \times V_\beta = V_{\alpha+\beta}}$,
\begin{equation}
\label{VabOPE}
V_\alpha(z)V_\beta(w) =(z-w)^{\alpha\beta}V_{\alpha+\beta}(w)+\ldots.
\end{equation}
Equipped with the bosonization rules, ${\mathcal{T}_k=V_{k/n}\bar{V}_{-k/n}}$ and ${\psi=V_1}$, the basic OPE that produces the desired $k$ component of the twist field from the fermion field is 
\begin{equation}
\psi \times V_{k/n-1} \bar{V}_{-k/n} = \mathcal{T}_k.
\end{equation}
Upon crossing the boundary line $x=0$ to $x<0$, the fermion field becomes the antiholomorphic fermion 
${\bar{\psi}=\bar{V}_1}$. We therefore also encounter the OPE whereby the fermion hits the antiholomorphic part of the twist field,
\begin{equation}
\bar\psi \times V_{k/n-1} \bar{V}_{-k/n} = \mathcal{T}_{k-n}.
\end{equation}
Since the leading singularities are ${(z-w)^{k/n-1}}$ and ${(z-\bar{w})^{-k/n}}$, we write a meromorphic function of the form
\begin{multline}\label{meroGCZ}
\langle \chi(z) V_{k/n-1}(w) \bar{V}_{-k/n}(\bar{w})  \rangle = \frac{1}{(z-w)^{1-k/n}(z-\bar{w})^{k/n}} \\
 \times \left(\frac{A(w,\bar{w})}{z-w}+\frac{\bar{A}(w,\bar{w})}{z-\bar{w}} +B(w,\bar{w}) \right).
\end{multline}
This has both the correct singular behaviour, and the appropriate decay at infinity of $z^{-1}$. For ${k=n/2}$, where the twist field obtains the scaling dimension of the spin field, this equation coincides with Eq.~(\ref{meroAB}).

One may expand this meromorphic function in powers of $z-w$, and compare with the OPE coefficients. We thereby obtain a closed set of coupled differential equations for both $\langle\mathcal{T}_{k}\rangle$ and $\langle\mathcal{T}_{k-n}\rangle$ as well as $\langle L_{-2}\mathcal{T}_{k}\rangle$ and $\langle L_{-2}\mathcal{T}_{k-n}\rangle$. As expected, these equations depend only on the dimensionless distance from the edge $\lambda\ell$. Moreover, by doing some algebraic manipulations,
these equations may be brought to the canonical form of a generalized hypergeometric ${}_2 F_3$ equation,

\begin{widetext}

\begin{equation}
\label{hg2f3}
\left\{\frac{d}{d\zeta}\left(\zeta\frac{d}{d\zeta}-\frac{1}{2}\right)^3-\left(\zeta\frac{d}{d\zeta}-\frac{1}{2}+\frac{k}{n}\right)\left(\zeta\frac{d}{d\zeta}-\frac{1}{2}-\frac{k}{n}\right)\right\}e^{-2\zeta^{1/2}}\mathfrak{t}_{k/n}(\zeta^{1/2})=0,
\end{equation}
with ${\zeta=\lambda^2\ell^2}$; for derivation see Appendix~\ref{app:gcz}.
This equation has a unique solution~\cite{hgpfq} satisfying the boundary conditions, ${\mathfrak{t}_{k/n}|_{\lambda\ell\to\infty}=1}$, which is
\begin{equation}
\label{fres}
\mathfrak{t}_{k/n}(\lambda\ell)=\frac{e^{2\lambda\ell}}{\pi^{5/2}}\left\{\pi^2 G_{2,4}^{2,2}\left(
\begin{array}{c}
\frac{1}{2}+\frac{k}{n}\,,\,\frac{1}{2}-\frac{k}{n}\\\noalign{\smallskip}
0\,,\,\frac{1}{2}\,,\,\frac{1}{2}\,,\,\frac{1}{2}
\end{array}\middle|\lambda^2\ell^2
\right)-\sin^2(\tfrac{\pi k}{n})G_{2,4}^{4,2}\left(
\begin{array}{c}
\frac{1}{2}+\frac{k}{n}\,,\,\frac{1}{2}-\frac{k}{n}\\\noalign{\smallskip}
0\,,\,\frac{1}{2}\,,\,\frac{1}{2}\,,\,\frac{1}{2}
\end{array}\middle|\lambda^2\ell^2
\right)\right\},
\end{equation}
where $G_{p,q}^{m,n}\left(
\substack{a_1,\ldots,a_p\\b_1,\ldots,b_q}\middle|z\right)$ is the Meijer G function. The properties and consequences of this solution is discussed in detail in the following section.

\end{widetext}

\section{Analytic results}
Eq.~(\ref{fres}) is an analytic universal result for the entanglemenet entropy. We now discuss the properties of the solution and compare with earlier numerics on lattice models. In the last section we will use it to predict the behavior of other models. 	   	  	   	  
Plugging into Eq.(\ref{BEdef}), we find an analytic expression for the R\'{e}nyi entropy,
\begin{equation}
\label{Renyi}
\mathfrak{s}_n(\ell)=\frac{1}{1-n}\ln\!\!\!\prod_{k>0}^{(n-1)/2} \!\!\!\mathfrak{t}_{k/n}(\lambda\ell).
\end{equation}
The function $\mathfrak{t}_{k/n}$ satisfies ${\mathfrak{t}_{0}(\lambda\ell)=1}$ as it should, since $\mathcal{T}_0$ is the identity field. It also possesses the nice property,
${\mathfrak{t}_{k/n}(0)=\cos(\pi k/n)}$,
which, using the trigonometric identity, ${\prod_{k>0}^{(n-1)/2}\cos\left(\frac{\pi k}{n}\right)=2^{-(n-1)/2}}$, leads to the remarkable conclusion,
\begin{equation}
\mathfrak{s}_n(0)=\frac{1}{1-n}\ln\!\!\!\prod_{k>0}^{(n-1)/2}\!\!\!\cos\left(\frac{\pi k}{n}\right)=\frac{1}{2}\ln 2,
\end{equation}
capturing analytically the half-qubit entropy for any $n$.
	   	  	   	  
So far, the parameter $k$ was treated as either integer or half-integer. Below, we consider two opposite limits by employing the analytic structure of the obtained functions in Eq.~(\ref{fres}), treating ${\beta=k/n}$ as a real number.
	   	  	   	  
The min-entropy is a relatively simple limit, $n\to\infty$, which directly follows from the definition of the Riemann integral, ${\mathfrak{s}_\infty(\ell)=-\int_0^{1/2}\ln(\mathfrak{t}_{\beta}(\lambda\ell))d\beta}$.   	  	   	  

\begin{figure}[t]
\centering
\includegraphics[width=1\linewidth]{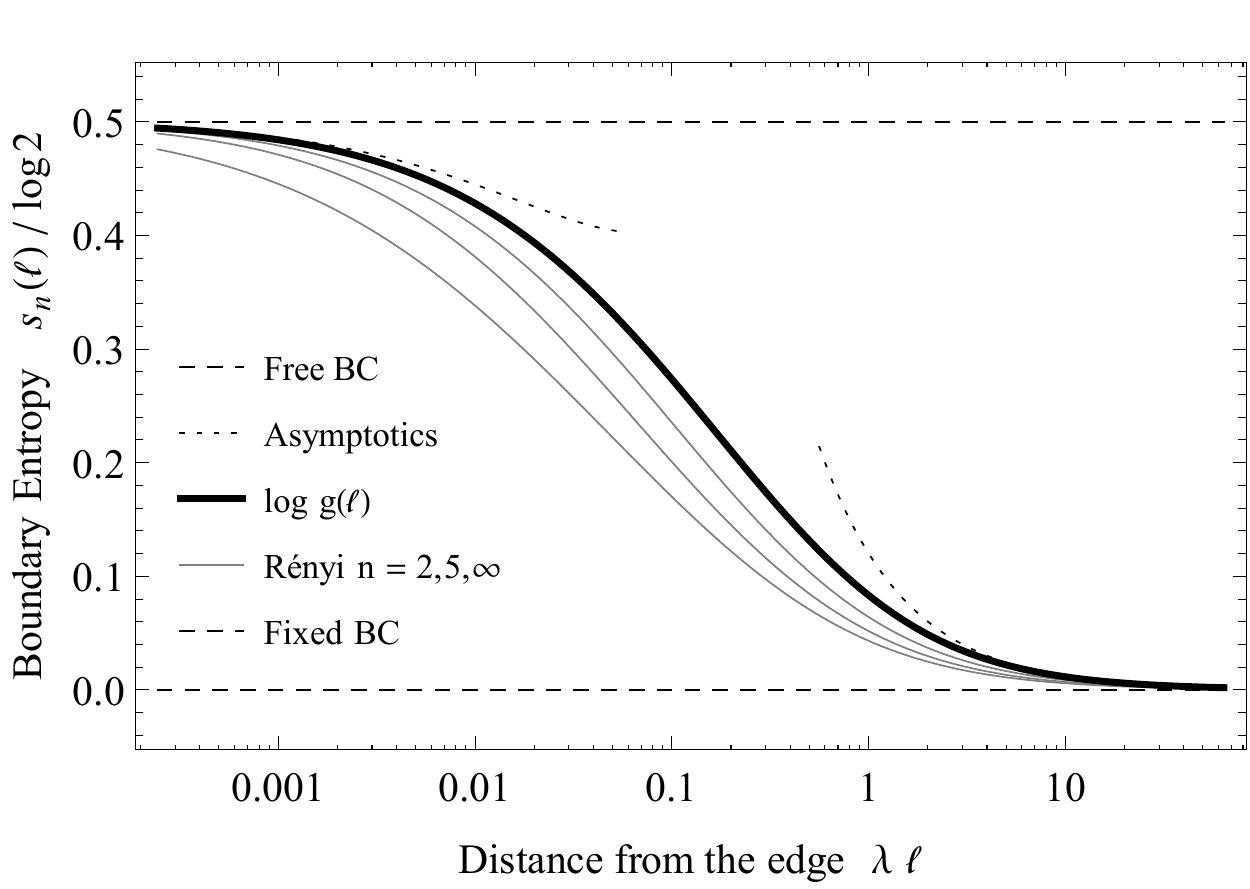}
\caption{Exact result for the universal RG flow of the boundary term of the entanglement entropy $\mathfrak{s}_n(\ell)/\log 2$ in the boundary Ising theory. Plotted are the boundary R\'{e}nyi entropies in Eq.~(\ref{Renyi}) for ${n=2,5,\infty}$, the groundstate degeneracy $g(\ell)$ limit $n \to 1$ in Eq.~(\ref{n0}), and its asymptotics in Eq.~(\ref{asymp}).}
\label{fig:FullRes}
\end{figure}

The main analytical result of this section is the groundstate degeneracy, $g(\ell)$, which corresponds to the limit of $n\to1$. Using a useful summation lemma proved in Appendix~\ref{app:lemma},
which follows from analyticity~\cite{levin1996lectures} and the Euler-Maclaurin formula, we arrive at the following expression for the boundary entanglement entropy,
\begin{equation}
\label{n0}
\mathfrak{s}_1(\ell)=\log[g(\ell)]=\sum_{\beta>0:\mathfrak{t}_{\beta}(\lambda\ell)=0}\left(\ln(\beta)-\psi^{(0)}(\beta)-\frac{1}{2\beta}\right).
\end{equation}
Here, ${\psi^{(0)}(z)=\Gamma'(z)/\Gamma(z)}$ is the Digamma function.
This expression can be computed by finding the zeros $\{ \beta \}$ of the function $\mathfrak{t}_{\beta}$. 
The analytic nature of this result allows one to study various entanglement properties such as its asymptotic behaviours,
\begin{equation}
\label{asymp}
\begin{aligned}
&\mathfrak{s}_1(\ell)\thicksim \frac{1}{2}\ln(2)-\frac{1}{4}\lambda\ell\ln^2(\lambda\ell), & \lambda\ell & \to 0,\\
&\mathfrak{s}_1(\ell)\thicksim \frac{1}{12\lambda\ell},  & \lambda\ell  & \to \infty.
\end{aligned}
\end{equation}
Details and further subleading asymptotics are given in Appendix~\ref{app:asymp}.
Indeed, at short distances one can apply perturbation theory with respect to free boundary conditions, ${h=0}$. The leading term arises from second order perturbation theory, which involves the correlation function $\int d t_1 \int dt_2  \langle \mathcal{T}(w,\bar{w}) \psi(t_1) a(t_1) \psi(t_2) a(t_2) \rangle $. From dimensional analysis, one arrives at a linear $\ell$ dependence, consistent with Eq.~(\ref{asymp}), which contains additional logarithmic corrections.
Both the form-factors approach~\cite{castro2009bi,saleur2013entanglement} and the free fermion solution~\cite{Casini2016} have demonstrated notable accuracy, however, neither have captured the precise form of this asymptotic behaviour.
This emphasizes the powerful analytic structure of the present solution. At long distances, the system is near the fixed boundary condition, ${h=\infty}$. The leading boundary irrelevant operator in the boundary Ising model is known to take the form $(\psi \partial_z \psi)_{x=0}$ with scaling dimension 2. This dictates~\cite{eriksson2011corrections} that the EE decays as $\ell^{-1}$ consistent with Eq.~(\ref{asymp}). All the results for $\mathfrak{s}_n(\ell)$ are plotted in Fig.~\ref{fig:FullRes}.
   	  	   	  
\begin{figure}[t]
\centering	
\includegraphics[width=1\linewidth]{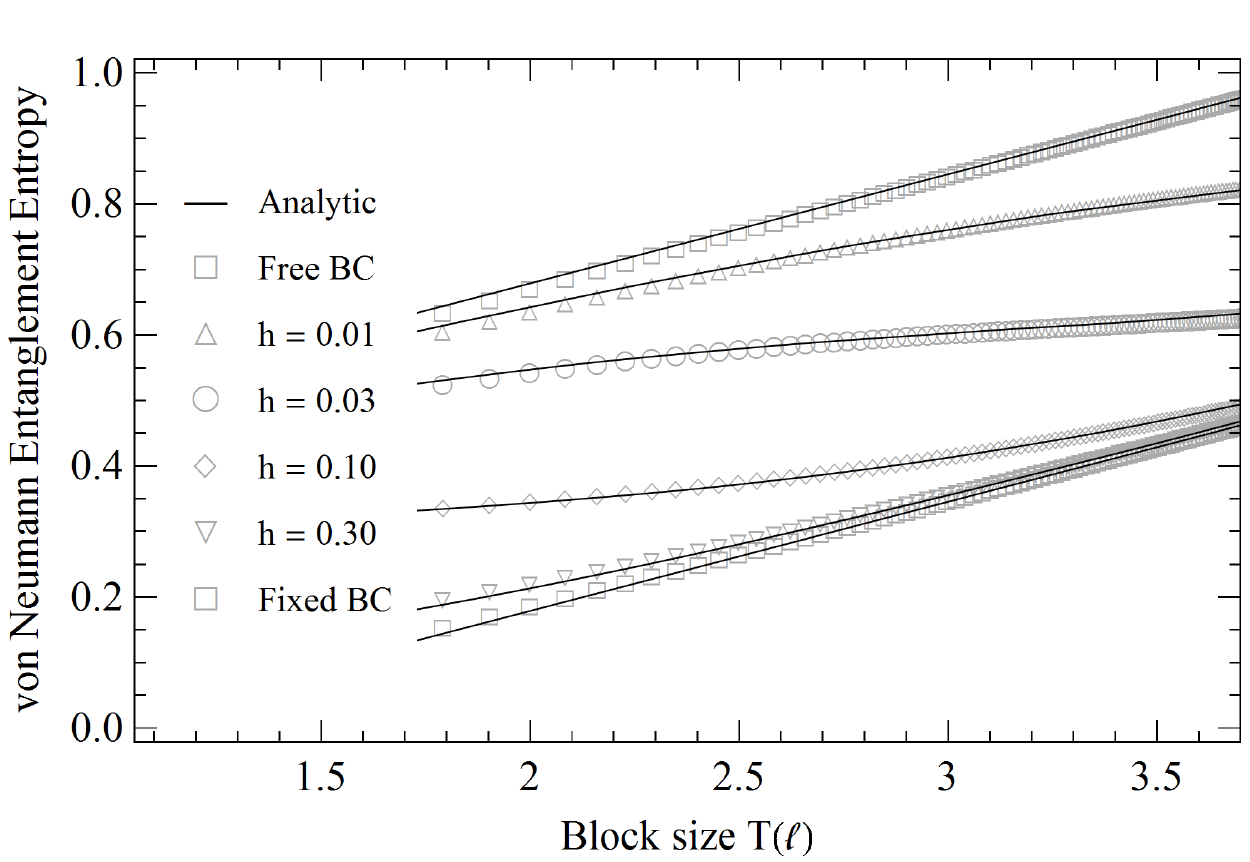}
\caption{Base 2 von Neumann Entanglement entropy $S_1(\ell)$ as function of the block size ${T(\ell)=\frac{1}{2} \log_2\left(\frac{2L}{\pi}\sin \left(\frac{\pi  \ell}{L}\right)\right)}$; data taken with permission from the work of Zhou \emph{et. al.}~\cite{zhou2006entanglement} is fitted to analytic result Eq.~(\ref{n0}).}
\label{fig:Comp}
\end{figure}

Interestingly, earlier attempts have been made to numerically tackle this crossover in the boundary term in the EE. Zhou \emph{et. al.}~\cite{zhou2006entanglement} computed via DMRG the EE in the boundary Ising chain Eq.~(\ref{bim}) for ${L=800}$ sites for various values of magnetic field $h_b$ applied at both boundaries [second boundary not included in Eq.~(\ref{bim})]. Our field theory calculation is restricted to a semi-infinite 1D system. It thus should describe the long distance physics solely in the limit, where the entanglement cut is far from the second boundary compared to the correlation length $\xi$. This corresponds to the numerical data $S(\ell)$ for $1 \ll \ell \ll L$ [in units where the lattice constant ${a=1}$]. We compare in Fig.~\ref{fig:Comp} the numerical data $S(\ell)$ for $5 \le \ell \le 90$. All points are fitted to Eq.~(\ref{n0}) with two fitting parameters: (i) A nonuniversal constant shift of the EE. This constant is fixed from fitting the the curves at ${h=0}$ and ${h= \infty}$. (ii) A regularization constant relating the magnetic field $h_b$ in the lattice model Eq.~(\ref{bim}) and $h$ in the field theory Eq.~(\ref{Sh}). We can see that for any given $h$ (or $h_b$), in this restricted regime  $1 \ll \ell \ll L$ where comparison to field theory is possible, the numerical data does not provide a full crossover from free to fixed boundary condition. This is due to the finite size of the studied system (${L=800}$). However, we can see that remarkably our single universal function fit all the numerical points which show this entire crossover upon incrasing $h$. We note that similar methods~\cite{leclair1996exact,sela2012local} could be used in the future to extend the field theory results to finite temperature or finite systems with two boundaries, as was simulated numerically.
	   	  	   	  
We have confirmed that Eq.~(\ref{n0}) is equivalent to half the corresponding result in Ref.~[\onlinecite{Casini2016}] which studied a field theory of the form Eq.~(\ref{Sh}) but with Dirac fermions instead of Majorana fermions and computed $\mathfrak{s}_1(\ell)$ using free fermion methods. Our methods are based on conformal symmetry and hence should have generalizations beyond the free fermion case.

%%%%

\section{Conclusions and Lookout}\label{conclusions}
In this article we have explored the scaling of the entanglemenet entropy of critical 1D systems with boundaries. Nontrivial renormalization group flow at the boundary is expressed in a universal scaling function in the entanglement entropy. We have combined CFT techniques~\cite{dixon1987conformal} in order to arrive at an analytic form of this universal crossover function for the boundary Ising model.

Previous results on the entanglement of two-channel Kondo models successfully identified the quenching of the entropy of the spin-$1/2$ impurity from $\log(2)$ down to $\frac{1}{2} \log(2)$~\cite{lee2015macroscopic}. Specifically, this universal function was extracted numerically~\cite{alkurtass2016entanglement} on a length scale $\xi_K$ - the Kondo screening cloud~\cite{affleck2001detecting,sorensen2007impurity,mitchell2011real,park2013directly}. It will be interesting to explore in the future the EE at the vicinity of the two-channel Kondo quantum critical point. For this purpose we propose to either (i) apply a small channel anisotropy, or (ii) apply a magnetic field at the $SU(2)$ symmetric impurity spin. Any combination of these perturbations quenches the ground state degeneracy. We predict that as long as these perturbations are small, the quenching of the residual half-qubit entropy occurs at a larger length scale $\xi \gg \xi_K$, and is described by the universal formula obtained here.

Our analytic results may be extended with further elaboration to a number of directions. This includes extensions to finite temperature, following Refs.~[\onlinecite{leclair1996exact},\onlinecite{sela2012local}] which extended the method of Chatterjee and Zamolodchikov~\cite{chatterjee1994local} by changing geometry to a semi-infinite cylinder; similarly, one may tackle finite size systems. The method 
can also be tested in excited states~\cite{Alcaraz2011,Taddia2016}. The method is fully analytic and allows the treatment of other measures of entanglement such as negativity~\cite{calabrese2012entanglement,calabrese2013entanglement,calabrese2013entanglement1,HOOGEVEEN2015,Calabrese2015Finite} corresponding to an analytic continuation to $n \to 1/2$.  An interesting question is whether additional critical theories beyond the Ising model can be treated using the methods described here.

\section{Acknowledgements}  	  
ES acknowledges Heung-Sun~Sim for his kind hospitality and discussions at KAIST in 2016 where this project started. EC acknowledges the Boulder school for providing CFT pedagogical background.	   	  	   	  	   	  	   	  
We thank Nissan Itzhaki, Zohar Komargodski, Jacob Sonnenschein, and particularly Pasquale Calabrese for useful comments and discussions, as well as Thomas Barthel and Ulrich Schollw\"{o}ck for kindly sharing with us  their data~\cite{zhou2006entanglement}.
This project was supported by an ISF Grant No. 1243/13,
and by a Marie Curie CIG Grant No. 618188.

\pagebreak
\onecolumngrid
\appendix
\section{Three point function}\label{app:3pf}
\renewcommand{\theequation}{A\arabic{equation}}
\setcounter{equation}{0}

In this appendix we derive the three point function in Eq.~(\ref{X}) using standard boundary CFT methods. 

Using the image method~\cite{cardy1984conformal} for conformal boundary conditions, we convert the correlators on the half plane $\mathcal{R}$ to correlators on the complex plane $\mathbb{C}$,
\begin{equation}
\langle \psi_{k}^*(z)\psi_{k}^{\phantom{*}}(z')\rangle_{\mathcal{R}_n}
=\frac{\langle \psi_{k}^*(z)\psi_{k}^{\phantom{*}}(z') \mathcal{T}(w,\bar{w})\rangle_{\mathcal{R}^n}}{\langle\mathcal{T}(w,\bar{w}) \rangle_{\mathcal{R}^n}}
=\frac{\langle \psi_{k}^*(z)\psi_{k}^{\phantom{*}}(z') \mathcal{T}(w) \tilde{\mathcal{T}}(\bar{w}) \rangle_{\mathbb{C}^n}}{\langle \mathcal{T}(w) \tilde{\mathcal{T}}(\bar{w}) \rangle_{\mathbb{C}^n}},
\end{equation}
where ${\mathcal{T}}(w)$ is the holomorphic part of ${\mathcal{T}}(w,\bar{w})$, and $\tilde{\mathcal{T}}(w)$ is the holomorphic part of the anti-twist field~\cite{calabrese2004entanglement} rotating in the opposite direction. The calculation of correlators in ${\mathbb{C}^n}$ can be performed using the uniformizing conformal transformation
\begin{equation}
\xi_j(z)=\xi(z)e_j=\left(\frac{z-w}{z-\bar{w}} \right)^{1/n}e^{2\pi i j/n}.
\end{equation}
It maps the $j$-th copy the complex plane in ${\mathbb{C}^n}$ into a wedge of angle $2 \pi / n$ in $\mathbb{C}$,
\begin{multline}
\frac{\langle \psi_{k}^*(z)\psi_{k}^{\phantom{*}}(z') \mathcal{T}(w) \tilde{\mathcal{T}}(\bar{w}) \rangle_{\mathbb{C}^n}}{\langle \mathcal{T}(w) \tilde{\mathcal{T}}(\bar{w}) \rangle_{\mathbb{C}^n}}
=\frac{1}{n}\sum_{jj'} e_j^{-k+n/2}e_{j'}^{k-n/2}\frac{\langle \psi_{j}^*(z)\psi_{j'}^{\phantom{*}}(z') \mathcal{T}(w) \tilde{\mathcal{T}}(\bar{w}) \rangle_{\mathbb{C}^n}}{\langle \mathcal{T}(w) \tilde{\mathcal{T}}(\bar{w}) \rangle_{\mathbb{C}^n}}\\
=\frac{1}{n}\sum_{jj'} e_j^{-k+n/2}e_{j'}^{k-n/2}(\xi'_{j}(z)\xi'_{j'}(z'))^{1/2}\langle \psi_{j}^*(\xi_j(z))\psi_{j'}^{\phantom{*}}(\xi_{j'}(z'))\rangle_{\mathbb{C}}
=\frac{1}{n}\sum_{jj'} e_j^{-k+n/2}e_{j'}^{k-n/2}\frac{(e_j\xi'(z)e_{j'}\xi'(z'))^{1/2}}{e_j\xi(z)-e_{j'}\xi(z')}.
\end{multline}
To evaluate the Fourier transform we expand to a power series and then resum the resulting expression,
\begin{multline}
\frac{1}{n}\sum_{jj'} e_j^{-k+n/2}e_{j'}^{k-n/2}\frac{(e_j\xi'(z)e_{j'}\xi'(z'))^{1/2}}{e_j\xi(z)-e_{j'}\xi(z')}
=\frac{1}{n}\left(\frac{\xi'(z)\xi'(z')}{\xi(z)^2}\right)^{1/2}\sum_{jj'}\sum_{p=0}^\infty e_j^{-k+(n-1)/2}e_{j'}^{k-(n-1)/2}\left(\frac{e_{j'}\xi(z')}{e_j\xi(z)}\right)^{p}\\
=n\left(\frac{\xi'(z)\xi'(z')}{\xi(z)\xi(z')}\right)^{1/2}\left(\frac{\xi(z')}{\xi(z)}\right)^{1/2}\sum_{q=0}^\infty \left(\frac{\xi(z')}{\xi(z)}\right)^{nq-k+(n-1)/2}\\
=n\left(\frac{\xi'(z)\xi'(z')}{\xi(z)\xi(z')}\right)^{1/2}\frac{\xi(z)^{k+n/2}\xi(z)^{-k+n/2}}{\xi(z)^n-\xi(z')^n}
=\frac{1}{z-z'}\left(\frac{(z-w)(z'-\bar{w})}{(z-\bar{w})(z'-w)} \right)^{k/n}
\end{multline}
For non-conformal finite $\lambda$, this is expected to hold far from the boundary ${\mathrm{Im}(w)=0}$ where the correlators are only sensitive to the bulk properties, i.e. for ${|z-z'|}, {|z-w|}, {|z-\bar{w}|}, {|z'-w|}, {|z'-\bar{w}|} \ll {|w-\bar{w}|}$.

\section{Generalized CZ equations}\label{app:gcz}
\renewcommand{\theequation}{B\arabic{equation}}
\setcounter{equation}{0}

In this appendix we show and solve the complete set of differential equations for the R\'{e}nyi boundary entropies that follow from Sec.~\ref{GCZ}.

We expand the meromorphic function of Eq.~(\ref{meroGCZ}) 
in powers of $z-w$ and $z-\bar{w}$, and compare with the known OPE coefficients of vertex operators in the free-boson theory,
\begin{multline}
\label{OPEvertex}
V_\alpha(z) V_\beta(w) = (z-w)^{\alpha \beta}\bigg(1+\frac{\alpha}{\alpha+\beta} (z-w) L_{-1}+(z-w)^2 \frac{\alpha \beta}{2 (\alpha+\beta)^2-1} L_{-2} \\ +(z-w)^2 \frac{\alpha (2 \alpha (\alpha+\beta)-1)}{2(\alpha+\beta)(2 (\alpha+\beta)^2-1)} L_{-1}^2+\ldots\bigg) V_{\alpha+\beta}(w).
\end{multline}
We thus obtain a closed set of coupled differential equations for both $\mathcal{T}_{k}$, $\mathcal{T}_{k-n}$ (and $L_{-2}$ acting on them). These lengthy equations are given by: 
\begin{eqnarray}
A&=&-\frac{(n-k) (w-\bar{w}) ^{k/n} {\langle \mathcal{T}_{k}\rangle}}{n},
\\
B+\frac{\bar{A}}{w-\bar{w} }&=&-\frac{(w-\bar{w}) ^{\frac{k}{n}-1}  \left(k n-k^2-i \lambda  n^2 (w-\bar{w}) \right)}{n^2}{\langle \mathcal{T}_{k}\rangle}+(w-\bar{w}) ^{k/n} {\partial_w\langle \mathcal{T}_{k}\rangle},
\\\nonumber
\bar{A}&=&-\frac{n (n-2 k) (k+n) (w-\bar{w}) ^{k/n+2} }{2k(n^2-2k^2)}{\partial_w^2\langle \mathcal{T}_{k}\rangle}-\frac{(w-\bar{w}) ^{\frac{k}{n}+1}  \left(k^2+i \lambda  n^2 (w-\bar{w}) \right)}{k n}{\partial_w\langle \mathcal{T}_{k}\rangle}\\
&&-\frac{k (w-\bar{w}) ^{\frac{k}{n}}  \left((k-n)^2+2 i \lambda  n^2 (w-\bar{w}) \right)}{2 n^3}{\langle \mathcal{T}_{k}\rangle}-\frac{\left(n^2-k^2\right) (w-\bar{w}) ^{k/n+2} }{n^2-2 k^2}{\langle L_{-2}\mathcal{T}_{k}\rangle},
\\
\bar{A}&=&\frac{k (w-\bar{w}) ^{1-\frac{k}{n}} {\langle \mathcal{T}_{k-n}\rangle}}{n},
\\
B-\frac{A}{w-\bar{w} }&=&-\frac{(w-\bar{w}) ^{-\frac{k}{n}}  \left(k n-k^2-i \lambda  n^2 (w-\bar{w}) \right)}{n^2}{\langle \mathcal{T}_{k-n}\rangle}+(w-\bar{w}) ^{1-\frac{k}{n}}{\partial_w\langle \mathcal{T}_{k-n}\rangle},
\\\nonumber
A&=&\frac{n (n-2 k) (2 n-k) (w-\bar{w}) ^{3-\frac{k}{n}} }{2 (n-k) \left(2 k^2-4 k n+n^2\right)}{\partial_w^2\langle \mathcal{T}_{k-n}\rangle}-\frac{(w-\bar{w}) ^{2-\frac{k}{n}}  \left((n-k)^2+i \lambda  n^2 (w-\bar{w}) \right)}{n (n-k)}{\partial_w\langle \mathcal{T}_{k-n}\rangle}\\
&&+\frac{(n-k) (w-\bar{w}) ^{1-\frac{k}{n}}  \left(k^2+2 i \lambda  n^2 (w-\bar{w}) \right)}{2 n^3}{\langle \mathcal{T}_{k-n}\rangle}-\frac{k (2 n-k) (w-\bar{w}) ^{3-\frac{k}{n}} }{2 k^2-4 k n+n^2}{\langle L_{-2}\mathcal{T}_{k-n}\rangle}.
\end{eqnarray}
Note that this is an infinite set of equations that involve all $k$'s. While the twist field is a product of $\mathcal{T}_k$ for a finite range of values of $k$ Eq.~(\ref{range}), 
the theory contains infinitely many such vertex operators as predicted by their OPE relation Eq.~(\ref{OPEvertex}). Luckily, these equations can be solved.
	
First, we algebraically solve for $A,\bar{A},B$; next, by shifting the index $k\to k+n$ in the last equation we can algebraically remove the dependence on ${\langle L_{-2}\mathcal{T}_{k-n}\rangle}$. We are then left with 2 differential equations for ${\langle \mathcal{T}_{k-n}\rangle},{\langle \mathcal{T}_{k}\rangle},{\langle \mathcal{T}_{k+n}\rangle}$.
To proceed, we define the ratio ${\beta = \frac{k}{n}}$ and the dimensionless distance 
\begin{equation}
X=-i(w-\bar{w})\lambda=2\lambda\ell.
\end{equation}
As a consequence of the above equations, one then gets that the normalized dimensionless twist, ${\mathfrak{t}_\beta(X) \propto X^{\beta^2}\langle \mathcal{T}_{k}\rangle}$, of Eq.(\ref{fdef}) satisfies
\begin{gather}
\beta^2(\mathfrak{t}_{\beta+1}+2\mathfrak{t}_{\beta}+\mathfrak{t}_{\beta-1})-2 X^2 \mathfrak{t}_\beta' + X^2 \mathfrak{t}_\beta''=0, \\ 
(1-2\beta)(\mathfrak{t}_{\beta}+\mathfrak{t}_{\beta-1})-X (\mathfrak{t}_{\beta}-\mathfrak{t}_{\beta-1})+X (\mathfrak{t}_{\beta}'-\mathfrak{t}_{\beta-1}')=0.
\end{gather}
Interestingly, one may form 3 closed equations by using the last equation for $\beta \to \beta+1$, giving
\begin{equation}
(1+2\beta)(\mathfrak{t}_{\beta}+\mathfrak{t}_{\beta+1})-X (\mathfrak{t}_{\beta}-\mathfrak{t}_{\beta+1})+X (\mathfrak{t}_{\beta}'-\mathfrak{t}_{\beta+1}')=0.
\end{equation}
This set of equations can be iterated to eliminate $\mathfrak{t}_{\beta\pm1}$ and yields the fourth order differential equation
\begin{eqnarray}
X^2 \mathfrak{t}_\beta''''+\left(3 X-4 X^2\right) \mathfrak{t}_\beta'''+\left(5 X^2-9 X+1\right) \mathfrak{t}_\beta''
+\left(-2 X^2+6 X-2\right) \mathfrak{t}_\beta'+4 \beta ^2 \mathfrak{t}_\beta=0.
\end{eqnarray}
This equation can be identified after exchanging ${X=2\sqrt{\zeta}}$, and ${\mathfrak{t}_\beta(X)=e^{X}f_\beta(\zeta)}$, whereby
\begin{equation}
\left\{\frac{d}{d\zeta}\left(\zeta\frac{d}{d\zeta}-\frac{1}{2}\right)^3-\left(\zeta\frac{d}{d\zeta}-\frac{1}{2}+\beta\right)\left(\zeta\frac{d}{d\zeta}-\frac{1}{2}-\beta\right)\right\}f_\beta(\zeta)=0.
\end{equation}
This is the canonical generalized hypergeometric ${}_2F_3$ equation of Eq.~(\ref{hg2f3}).

\section{A summation Lemma}\label{app:lemma}
\renewcommand{\theequation}{C\arabic{equation}}
\setcounter{equation}{0}

We present here the proof and application of a useful summation lemma which is essential for the calculation of the groundstate degeneracy in Eq.(\ref{n0}). 

\subsection{Lemma}

let $h(z):\mathbb{C}\to\mathbb{C}$ be an analytic function of $z$ such that:
\begin{itemize}
	\item $h(z)$ is an entire function of order strictly lesser than 2,
	\item $h(\mathbb{R})\subseteq\mathbb{R}$,
	\item $h(z)=h(-z)$,
	\item $h(0)=1$,
	\item $\{z:h(z)=0\}\subset\mathbb{R}$,
	\item $\forall z\in(-\tfrac{1}{2},\tfrac{1}{2}),~h(z)>0$.
\end{itemize}
Under these conditions
\begin{equation}
\Delta[h]\equiv-\lim_{m\to0}\frac{1}{2m}\ln\prod_{k=0}^{m}h(\tfrac{k}{2m+1})=\sum_{z>0:h(z)=0}N_z\cdot\left(\ln(z)-\psi^{(0)}(z)-\frac{1}{2z}\right),
\end{equation}
where $N_z$ is the multiplicity of the zero $z$, and ${\psi^{(0)}(z)=\Gamma'(z)/\Gamma(z)}$ is the Digamma function.

\subsection{Proof}

We start by using the Euler–Maclaurin formula for a function $f(k,m)$,
\begin{equation}
\sum_{k=0}^m f(k,m)=\int\limits_0^m f(k,m)dk+\frac{f(m,m)+f(0,m)}{2}+\sum_{p=1}^\infty \frac{B_{2p}}{(2p)!}\left[\partial_k^{2p-1}f(k,m)\right]_{k=0}^{k=m},
\end{equation}
where $B_p$ are the Bernoulli numbers.
By taking the derivative with respect to $m$, and reusing the Euler-Maclaurin formula for $\partial_k f(k,m)$ and $\partial_m f(k,m)$, one gets
\begin{equation}
\partial_m\sum_{k=0}^m f(k,m)=\sum_{k=0}^m (\partial_k+\partial_m)f(k,m)+\sum_{p=0}^\infty \frac{B_p}{p!}\left.\partial_k^{p}f(k,m)\right|_{k=0}.
\end{equation}
By setting ${f(k,m)=\ln(h(\tfrac{k}{2m+1}))}$, we can use the properties ${h(0)=1}$ and ${\partial_z^{2p-1}h(z)|_{z=0}=0}$ to get,
\begin{equation}
\Delta[h]=-\frac{1}{2}\lim_{m\to0}\partial_m\sum_{k=0}^{m}\ln(h(\tfrac{k}{2m+1}))
=-\frac{1}{2}\sum_{p=1}^\infty \frac{B_{2p}}{(2p)!}\left.\partial_z^{2p}\ln(h(z))\right|_{z=0}.
\end{equation}
To evaluate the derivatives we first use Cauchy's integral formula; next, since $h(z)$ is even, we may invert the integral, integrate over its positive zeros ${Z_0^+\equiv\{z>0:h(z)=0\}}$, and utilize the generalized argument principle,
\begin{equation}
\Delta[h]=-\frac{1}{2}\sum_{p=1}^\infty\frac{B_{2p}}{(2p)!}\frac{(2p-1)!}{2\pi i}\oint\limits_{z=0}\frac{h'(z)dz}{h(z)z^{2p}}=\oint\limits_{z\in Z_0^+}\frac{dz}{2\pi i}\sum_{p=1}^\infty\frac{B_{2p}}{2p z^{2p}}\frac{h'(z)}{h(z)}=\sum_{z\in Z_0^+}N_z\sum_{p=1}^\infty\frac{B_{2p}}{2p z^{2p}}.
\end{equation}
We now recall the asymptotic expansion of the Digamma function ${\psi^{(0)}(z)=\Gamma'(z)/\Gamma(z)}$,
\begin{equation}
\psi^{(0)}(z)=\ln(z)-\frac{1}{2z}-\sum_{p=1}^\infty\frac{B_{2p}}{2p z^{2p}}.
\end{equation}
And thus
\begin{equation}\label{finLemma}
\Delta[h]=\sum_{z>0:h(z)=0}N_z\cdot\left(\ln(z)-\psi^{(0)}(z)-\frac{1}{2z}\right).
\end{equation}
All that is left is to prove convergence of this sum.

We have demanded $h$ to be of an analytic function of order strictly lesser than 2, therefore, it is known~\cite{levin1996lectures} that the sum ${\sum_{z:h(z)=0}N_z|z|^{-2}}$ converges. Since for large zeros one has
\begin{equation}\label{asymp1}
\ln(z)-\psi^{(0)}(z)-\frac{1}{2z}=\frac{1}{12z^2}+O(z^{-4}),
\end{equation}
we immediately find that the sum in Eq.(\ref{finLemma}) is convergent.
\begin{flushright}
	Q.E.D.
\end{flushright}

\subsection{Application}

Using this lemma we may now evaluate the boundary entropy, $\mathfrak{s}_1(X)$, from Eq.~(\ref{Renyi}) by setting ${h(z)=\mathfrak{t}_z(X)}$ and ${n=2m+1}$. It is fairly straightforward to show that $\mathfrak{t}$ follows the conditions of the lemma. Therefore, since $\mathfrak{t}_z(X)$ has but simple zeros in $z$, we have
\begin{equation}
\mathfrak{s}_1(X)=-\lim_{n\to 1}\frac{1}{n-1}\sum_{k>0}^{(n-1)/2}\ln(\mathfrak{t}_{k/n}(X))=\sum_{\beta>0:\mathfrak{t}_\beta(X)=0}\left(\ln(\beta)-\psi^{(0)}(\beta)-\frac{1}{2\beta}\right).
\end{equation}

Much like the limit case, ${\mathfrak{t}_\beta(0)=\cos(\pi\beta)}$, the function $\mathfrak{t}_\beta(X)$ is entire of order 1 in $\beta$, and for every $X$ its zeros,  ${{\{\beta_j\}_{j=0}^{\infty}}={\{\beta>0:\mathfrak{t}_\beta(X)=0\}}}$, are simple and satisfy ${\beta_{j}=O(j)}$ for large $j$; \emph{e.g.} for ${X=0}$ one has ${\beta_j=j+\frac{1}{2}}$. Using Eq.(\ref{asymp1}), this property allows for the evaluation of the asymptotic behaviour of the summand and of the partial sums,
\begin{equation}
\ln(\beta_{j})-\psi^{(0)}(\beta_{j})-\frac{1}{2\beta_{j}}=O(j^{-2}),\qquad
\mathfrak{s}_1(X)-\sum_{j'=0}^j\left(\ln(\beta_{j'})-\psi^{(0)}(\beta_{j'})-\frac{1}{2\beta_{j'}}\right)=O(j^{-1}).
\end{equation}
This slow but very predictable asymptotic behaviour allows one to evaluate the infinite sum to a satisfying accuracy using a relatively small number of zeros, and makes its calculation very efficient using acceleration methods such as the rational function extrapolation.

\section{Asymptotic Expansion}\label{app:asymp}
\renewcommand{\theequation}{D\arabic{equation}}
\setcounter{equation}{0}
In this appendix we derive the asymptotic expansions of the groundstate degeneracy given in Eq.~(\ref{asymp}). 
These expansions, especially for large $X$, also provide a computationally efficient way to find $\mathfrak{s}_1(X)$.
\subsection{Long distance expansion}
Either by applying the Frobenius method over the ordinary differential equation of Eq.~(\ref{hg2f3}), 
or directly from the analytic expression for $\mathfrak{t}_\beta(X)$ of Eq.~(\ref{fres}), 
we first calculate the asymptotic power series,
\begin{equation}
-\ln(\mathfrak{t}_\beta(X))\thicksim \frac{2\beta^2}{X}-\frac{2\beta^2}{X^2}+\frac{\frac{10}{3}\beta^2+\frac{10}{3}\beta^4}{X^3}-\frac{8\beta^2+20\beta^4}{X^4}+\frac{\frac{128}{5}\beta^2+108\beta^4+\frac{84}{5}\beta^6}{X^5}+\ldots
\equiv\sum_{p=1}^\infty\sum_{q=1}^{\lceil p/2 \rceil} a_{p,q}\frac{\beta^{2q}}{X^p}.
\end{equation} 
Next, we may utilize the nice identity,
\begin{equation}
\lim_{n\to 1}\frac{1}{n-1}\sum_{k>0}^{(n-1)/2}\left(\frac{k}{n}\right)^{2q}=\lim_{n\to 1}\frac{1}{n-1}\cdot\frac{B_{2q+1}(\tfrac{n+1}{2})-B_{2q+1}(\tfrac{1-n}{2})}{2(2q+1)}=\frac{1}{2}B_{2q},
\end{equation}
where $B_q,B_q(x)$ are the Bernoulli numbers and polynomials. We therefore have
\begin{equation}
\mathfrak{s}_1(X)=-\lim_{n\to 1}\frac{1}{n-1}\sum_{k>0}^{(n-1)/2}\ln(\mathfrak{t}_{k/n}(X))\thicksim\sum_{p=1}^\infty\sum_{q=1}^{\lceil p/2 \rceil} a_{p,q}\frac{1}{2}B_{2q}\frac{1}{X^p}= \frac{1}{6X}-\frac{1}{6X^2}+\frac{2}{9X^3}-\frac{1}{3X^4}+\frac{8}{15X^5}+\ldots.
\end{equation}
Although this asymptotic series has a zero convergence radius and hence diverges for all $|X|<\infty$, it may nevertheless be summed using superasymptotics methods to a satisfying accuracy for large enough $X$. Specifically, for $X\gtrsim4$ its superasymptotic summation agrees with the exact results of Appendix~\ref{app:lemma} to within at least 3 significant digits.

\subsection{Short distance expansion}
The small $X$ expansion of $\mathfrak{s}_1(X)$ is done using a similar technique to that of the summation lemma.
We use the known properties of the Meijer G function to expand Eq.~(\ref{fres}) and get
\begin{equation}
\ln(\mathfrak{t}_\beta(X))\thicksim \ln(\cos(\pi\beta))+X\ln^2(X)\frac{2\beta}{\pi}\tan(\pi\beta)+\ldots.
\end{equation} 
By repeating the derivation of the lemma we find
\begin{equation}
\mathfrak{s}_1(X)\thicksim\frac{1}{2}\ln(2)-X\ln^2(X)\frac{1}{2}\sum_{p=1}^\infty \frac{B_{2p}}{(2p)!}\left.\partial_z^{2p}\frac{2z}{\pi}\tan(\pi z)\right|_{z=0}+\ldots.
\end{equation} 
To evaluate the derivatives, we first use Cauchy's integral formula; next, we may invert the integral, and use reflection symmetry to integrate over its positive poles ${z_j=\frac{1}{2}+j}$,
\begin{multline}
\frac{1}{2}\sum_{p=1}^\infty \frac{B_{2p}}{(2p)!}\left.\partial_z^{2p}\frac{2z}{\pi}\tan(\pi z)\right|_{z=0}=\frac{1}{2}\sum_{p=1}^\infty B_{2p}\oint\limits_{z=0}\frac{\frac{2z}{\pi}\tan(\pi z)}{z^{2p+1}}\frac{dz}{2\pi i}=\\
=-\sum_{p=1}^\infty\sum_{w=\frac{1}{2}}^\infty \frac{2B_{2p}}{\pi w^{2p}}\oint\limits_{z=w}\tan(\pi z)\frac{dz}{2\pi i}=\sum_{z=\frac{1}{2}}^\infty\sum_{p=1}^\infty \frac{2B_{2p}}{\pi^2 z^{2p}}=\sum_{z=\frac{1}{2}}^\infty\frac{2z}{\pi^2}\left(\psi^{(1)}(z)-\frac{1}{z}-\frac{1}{2z^2}\right)=\frac{1}{8}.
\end{multline}
Here, we have recalled the asymptotic expansion and summation properties of the first Polygamma function ${\psi^{(1)}(z)=\partial_z\psi^{(0)}(z)
=z^{-1}+\frac{1}{2}z^{-2}+\sum_{p=1}^\infty B_{2p}z^{-2p-1}}$.
We therefore conclude that
\begin{equation}
\mathfrak{s}_1(X)\thicksim\frac{1}{2}\ln(2)-\frac{1}{8}X\ln^2(X)+\ldots.
\end{equation}
Note that further coefficients may be found using similar techniques. 

\twocolumngrid
%\bibliographystyle{elsart-num}
%\bibliographystyle{apsrmp}
%\newpage
%\bibliographystyle{elsart-num}
\bibliographystyle{apsrev4-1}
%\bibliographystyle{apsrmp}
%\bibliographystyle{plain}
%\bibliography{EErefs}


\end{document}